\documentclass[aps,prd,twocolumn,showkeys,amsmath,amssymb]{revtex4}
\usepackage{graphicx}
\usepackage{epstopdf}
\usepackage{multirow}
\usepackage{subfigure}
\usepackage{extarrows}
\usepackage{feynmf}
\usepackage{enumitem}
\usepackage{color}
\usepackage[colorlinks,citecolor=blue,anchorcolor=red,menucolor=red,linkcolor=red,filecolor=red,runcolor=red,urlcolor=blue,frenchlinks=red]{hyperref}

\begin{document}

\title{New molecular bonds existing in the strong interaction}

\author{Hua-Xing Chen}
\email{hxchen@seu.edu.cn}
\affiliation{School of Physics, Southeast University, Nanjing 210094, China}

\begin{abstract}
Similar to the covalent bond in chemical molecules induced by shared electrons, we proposed in Ref.~\cite{Chen:2021xlu} the hadronic covalent bond induced by shared light quarks to explain the $T_{cc}(3875)$ and the deuteron. In this paper we improve and extend this mechanism to explain the $Z_c(3900)$, which is bound by the shared light quark-antiquark pair along with sea quark-antiquark pairs from the vacuum. Our analysis is based on the following forward and backward reasoning: a hadronic molecule exists, iff the attraction between its components is strong enough, iff the wave functions of its components significantly overlap with each other, iff the Pauli principle is well satisfied among all the shared quarks and antiquarks. Additionally, the $X(3872)$ is so unique that we need to further consider the annihilation of the shared light quark-antiquark pair, just in line with the reasoning that the creation and annihilation of sea quark-antiquark pairs should be given equal consideration. Both the ``creation'' and ``annihilation'' molecular bonds exist only in the strong interaction, not in the electromagnetic interaction, and they provide a quasi-static low-energy platform for studying the QCD confinement.
\end{abstract}

\date{\today}
\maketitle

\section{Introduction}
\label{sec:intro}

The past decade has marked a golden era in hadron physics~\cite{pdg}, which significantly improves our understanding of the non-perturbative behavior of the strong interaction in the low-energy regime~\cite{Chen:2016qju,Liu:2019zoy,Chen:2022asf,Hosaka:2016pey,Richard:2016eis,Lebed:2016hpi,Esposito:2016noz,Ali:2017jda,Guo:2017jvc,Olsen:2017bmm,Karliner:2017qhf,Guo:2019twa,Brambilla:2019esw,Meng:2022ozq,Liu:2024uxn}. This now presents an ideal opportunity to compare it with the electromagnetic interaction through the investigation of the hydrogen atom/molecule and the charmed meson/molecule.

A hydrogen atom consists of an electron bound to a proton, with a binding energy of 13.6~eV. In the ground state, the interaction between the electron and proton spins results in a slight increase in energy when the spins are parallel, and a decrease when antiparallel. The energy difference between these two hyperfine states is approximately $5.87 \times 10^{-6}$~eV. The situation become more complex in the hydrogen molecule, which consists of two hydrogen atoms bound together by a chemical covalent bond, with a binding energy of 4.73~eV. There are two spin configurations of the hydrogen molecule, {\it i.e.}, ortho-hydrogen has parallel proton spins and higher energy, while para-hydrogen has antiparallel proton spins and lower energy. The energy difference between these two spin isomers is approximately $0.0151$~eV.

A comparable picture has been found in the strong interaction, namely, the charmed meson and its hadronic molecule. A charmed meson consists of a light up/down antiquark bound to a charm quark within the conventional quark model. There are two ground states: \(D\) with \(J^P = 0^-\) and \(D^*\) with \(J^P = 1^-\), with a mass difference of about 140~MeV. For brevity, we denote them collectively as \(D^{(*)}\), and refer to \(\bar{D}\) and \(\bar{D}^*\) collectively as \(\bar{D}^{(*)}\). Two $\bar D^{(*)}$ mesons may form a $\bar D^{(*)}\bar D^{(*)}$ hadronic molecule, but so far only the $T_{cc}(3875)$ has been observed in the LHCb experiment~\cite{LHCb:2021auc,LHCb:2021vvq}, which can be interpreted as the $\bar D \bar D^*$ hadronic molecule of $(I)J^P = (0)1^+$, with a binding energy on the order of 1~MeV. More $\bar D^{(*)}\bar D^{(*)}$ hadronic molecules, such as the $\bar D^* \bar D^*$ hadronic molecule of $(I)J^P = (0)1^+$, have been predicted in the literature and are yet to be discovered, which would contribute to completing this picture~\cite{Chen:2016qju,Liu:2019zoy,Chen:2022asf,Hosaka:2016pey,Richard:2016eis,Lebed:2016hpi,Esposito:2016noz,Ali:2017jda,Guo:2017jvc,Olsen:2017bmm,Karliner:2017qhf,Guo:2019twa,Brambilla:2019esw,Meng:2022ozq,Liu:2024uxn}.

In addition to the many similarities between the hydrogen atom/molecule and the charmed meson/molecule, there are two distinct differences:
\begin{itemize}

\item In the hydrogen atom and the hydrogen molecule, the contribution from the proton spin is small and negligible. In contrast, in the $\bar D^{(*)}$ meson and its hadronic molecule, the contribution from the spin of the $charm$ antiquark is no longer negligible.

\item The binding energy of the hydrogen molecule, resulting from the residual electromagnetic interaction between two hydrogen atoms, is of the same order as that of the hydrogen atom, which arises from the direct electromagnetic interaction between the electron and proton. In contrast, the binding energy of the $T_{cc}(3875)$, resulting from the residual strong interaction between the $\bar D$ and $\bar D^*$ mesons, is significantly smaller than the ``binding'' energy of the $\bar D^{(*)}$ meson, which arises from the direct strong interaction between the light $up/down$ quark and the $charm$ antiquark.

\end{itemize}
Another related observation is that a muonic $Z$ atom is believed to behave as if it were a $Z-1$ atom, though this will not be discussed in the present study.

%
\begin{figure*}[htbp]
\begin{center}
\subfigure[~Covalent bond: $T_{cc}(3875)$]{\includegraphics[width=0.31\textwidth]{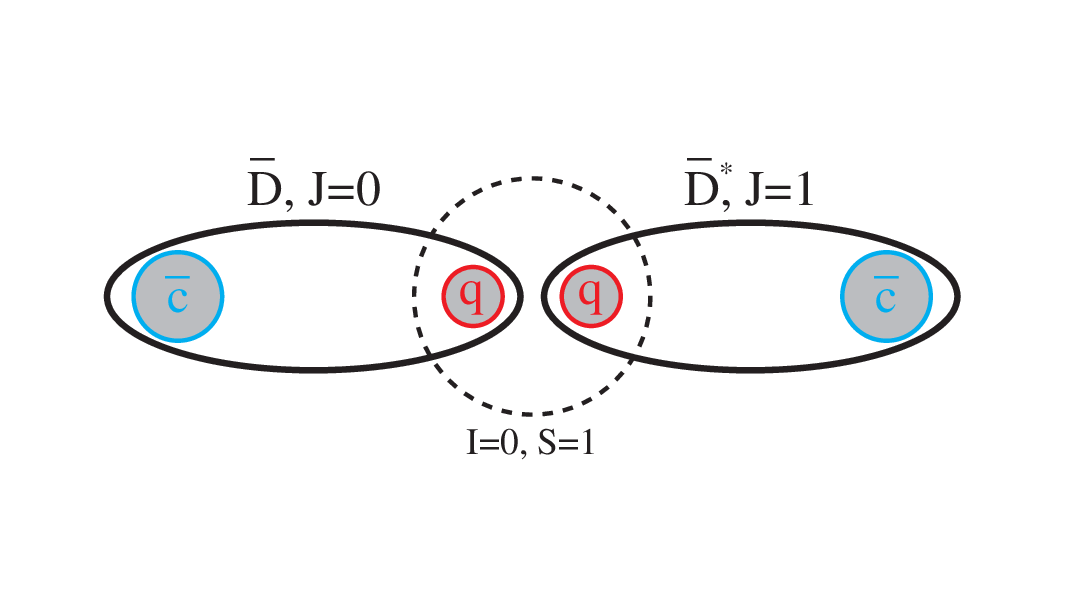}}
~~~
\subfigure[~Creation bond: $Z_c(3900)$]{\includegraphics[width=0.31\textwidth]{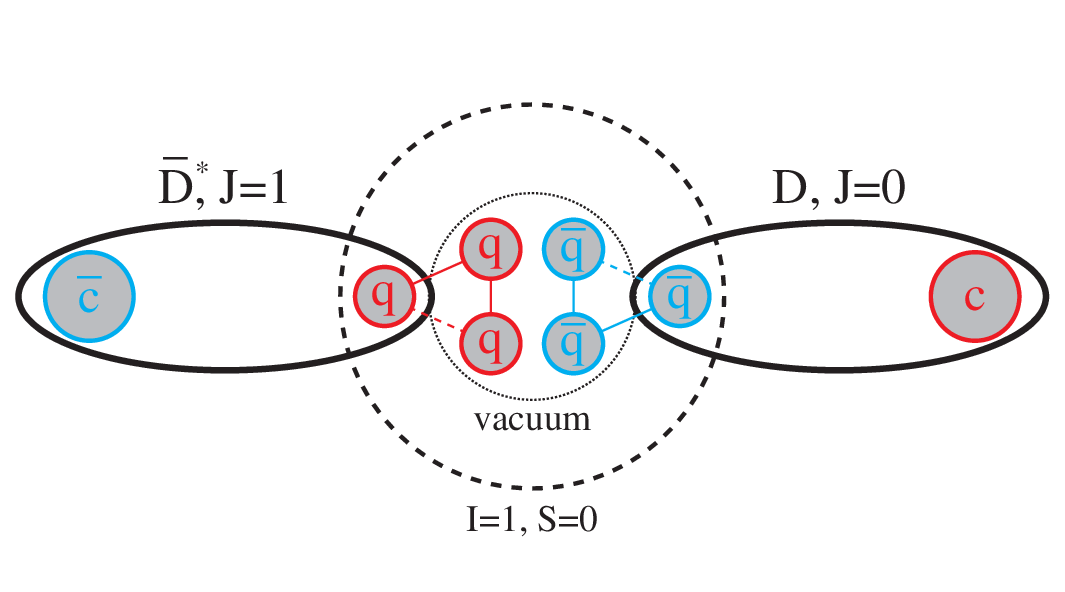}}
~~~
\subfigure[~Annihilation bond: $X(3872)$]{\includegraphics[width=0.31\textwidth]{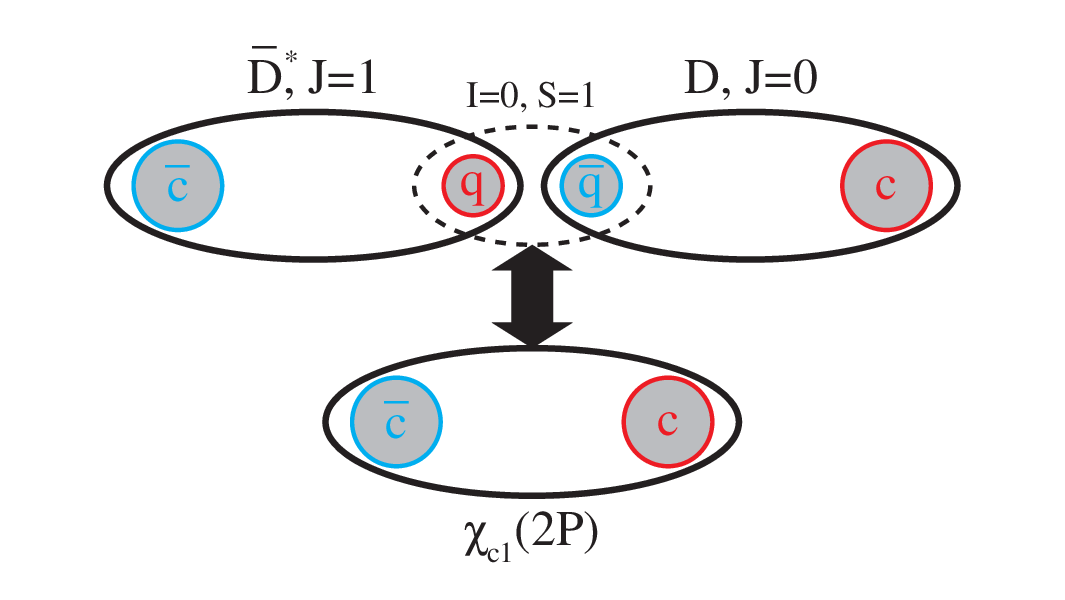}}
\caption{Possible binding mechanisms induced by: (a) shared light quarks, (b) the shared light quark-antiquark pair along with sea quark-antiquark pairs from the vacuum, and (c) the annihilation of the shared light quark-antiquark pair.}
\label{fig:bond}
\end{center}
\end{figure*}
%

Similar to the chemical covalent bond, we proposed in Ref.~\cite{Chen:2021xlu} the hadronic covalent bond to explain the $P_c/P_{cs}/T_{cc}$~\cite{LHCb:2015yax,LHCb:2019kea,LHCb:2021chn,LHCb:2020jpq,LHCb:2022ogu,LHCb:2021vvq,LHCb:2021auc} and the deuteron as possible hadronic covalent molecules. In the hydrogen molecule, the two electrons shared between two protons are antisymmetric so that obey the Pauli principle. Similarly, we proposed in the $T_{cc}(3875)$ that the two light $up/down$ quarks shared between two $charm$ antiquarks are also antisymmetric so that obey the Pauli principle, as illustrated in Fig.~\ref{fig:bond}(a). Since the binding energy of the $T_{cc}(3875)$ is significantly smaller than the ``binding'' energy of the $\bar D^{(*)}$ meson, the two shared light $up/down$ quarks need not be fully antisymmetric, as long as there is enough antisymmetric component to provide sufficient attraction.

However, the hadronic covalent bond cannot explain the hadronic molecule composed of one $D^{(*)}$ meson and one $\bar D^{(*)}$ meson. Specifically, the $Z_c(3900)$~\cite{BESIII:2013ris,Belle:2013yex} and the $X(3872)$~\cite{Belle:2003nnu} can be interpreted as such hadronic molecules~\cite{Chen:2016qju,Liu:2019zoy,Chen:2022asf,Hosaka:2016pey,Richard:2016eis,Lebed:2016hpi,Esposito:2016noz,Ali:2017jda,Guo:2017jvc,Olsen:2017bmm,Karliner:2017qhf,Guo:2019twa,Brambilla:2019esw,Meng:2022ozq,Liu:2024uxn}. An interesting question related to the electromagnetic interaction is whether there exists the chemical molecule composed of one hydrogen atom and one anti-hydrogen atom. While the answer is not yet known to me, it is likely that additional binding mechanisms exist in the strong interaction.

In this paper we extend the hadronic covalent bond to explain the $Z_c(3900)$ as the $D \bar D^*/D^* \bar D$ hadronic molecule with $I^GJ^{PC} = 1^+1^{+-}$. As illustrated in Fig.~\ref{fig:bond}(b), we consider not only the light quark-antiquark pair from the $D$ and $\bar D^*$ mesons, but also two sea quark-antiquark pairs from the vacuum. Consequently, the annihilation of the light quark-antiquark pair should also be considered. As illustrated in Fig.~\ref{fig:bond}(c), this effect has been studied in Refs.~\cite{Meng:2005er,Meng:2014ota,Matheus:2009vq,Brambilla:2024thx} to explain the $X(3872)$ as the mixture of a $c \bar c$ state and the $D \bar D^*/D^* \bar D$ component with $I^GJ^{PC} = 0^+1^{++}$, and this analysis is closely related to the $^3P_0$ model studies on the hadronic decays of charmonium~\cite{Micu:1968mk,LeYaouanc:1972vsx,Barnes:2005pb}. We refer to the two binding mechanisms above as the ``creation'' and ``annihilation'' bonds, which exist only in the strong interaction, not in the electromagnetic interaction. These two bonds provide a quasi-static low-energy platform for studying the QCD confinement, so we refer to them collectively as the ``confined'' bond, with the relevant molecule termed the hadronic ``confined'' molecule.

This paper is organized as follows. In Sec.~\ref{sec:covalent} we improve the hadronic covalent bond proposed in Ref.~\cite{Chen:2021xlu}, and apply it to study the $T_{cc}(3875)$ and the deuteron. In Sec.~\ref{sec:creation} we propose the hadronic creation bond to explain the $Z_c(3900)$ through the shared light quark-antiquark pair along with sea quark-antiquark pairs from the vacuum. In Sec.~\ref{sec:summary} we use the hadronic annihilation bond to explain the $X(3872)$ by further considering the annihilation of the light quark-antiquark pair, after which we conclude the paper.

\section{Covalent bond}
\label{sec:covalent}

\begin{figure}[htbp]
\begin{center}
\includegraphics[width=0.4\textwidth]{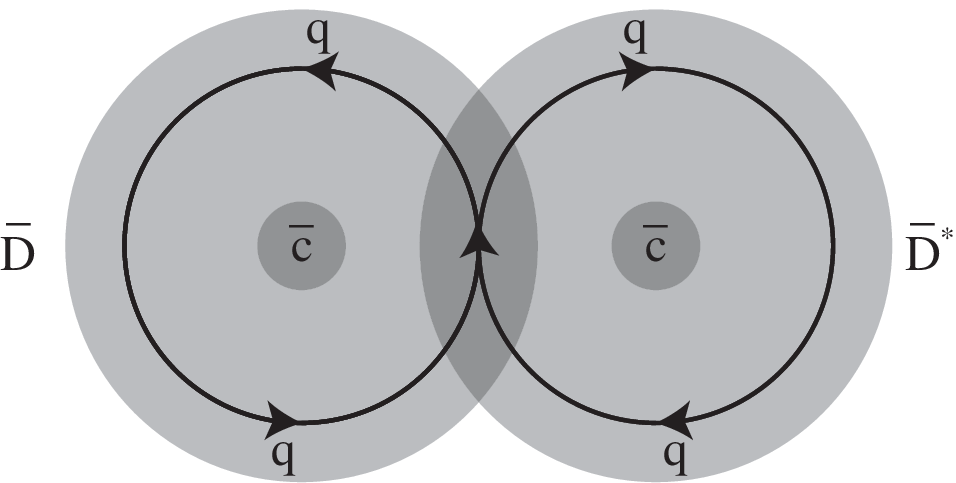}
\caption{Possible binding mechanism for the $T_{cc}(3875)$ as the $\bar D \bar D^*$ hadronic molecule with $(I)J^P = (0)1^+$, due to the hadronic covalent bond induced by shared light quarks.}
\label{fig:covalent}
\end{center}
\end{figure}

We proposed in Ref.~\cite{Chen:2021xlu} the hadronic covalent bond induced by shared light quarks, as illustrated in Fig.~\ref{fig:covalent}. In the chemical covalent bond the shared electrons are totally antisymmetric so that obey the Pauli principle. Similarly, we found in the hadronic covalent bond that ``{\it the light-quark-exchange interaction is attractive when the shared light quarks are totally antisymmetric so that obey the Pauli principle}''. In this section we improve this mechanism and apply it to study several examples in the following subsections.

%
%

\subsection{$T_{cc}(3875)$}
\label{sec:Tcc}

In this subsection we study the hadronic molecules composed of two $\bar D^{(*)}$ mesons. In order to form these molecules, the light quark from one $\bar D^{(*)}$ meson and the light quark from the other $\bar D^{(*)}$ meson should be totally antisymmetric according to the Pauli principle. This ensures that the wave functions of the two $\bar D^{(*)}$ mesons significantly overlap with each other, resulting in a strong attraction that may lead to the formation of a hadronic covalent molecule. The two light quarks shared between the two $\bar D^{(*)}$ mesons have the same color and so the symmetric color structure. Besides, we assume their orbital structure to be $S$-wave and so also symmetric. Consequently, we only need to investigate their spin and flavor structures ($q=up/down$):
\begin{itemize}

\item{$\bar D^0[u_1 \bar c_2]$--$\bar D^0[u_3 \bar c_4]$.} Let us exchange $u_1$ from the first $\bar D^0$ meson with $u_3$ from the second $\bar D^0$ meson. $u_1$ and $\bar c_2$ inside the first $\bar D^0$ meson spin in opposite directions, $u_3$ and $\bar c_2$ also need to spin in opposite directions in order to form another $\bar D^0$ meson, so $u_1$ and $u_3$ spin in the same direction with the symmetric spin structure. The flavor structure of $u_1$ and $u_3$ is also symmetric, so they are totally symmetric (${\bf S}$=symmetric and ${\bf A}$=antisymmetric):
\begin{equation}\nonumber
\begin{array}{cccccc}
\hline
                           & {\rm color} & {\rm flavor} & {\rm spin} & {\rm orbital} & {\rm total}
\\ \hline
u_1 \leftrightarrow u_3    & {\bf S}     & {\bf S}      & {\bf S}    & {\bf S}       & {\bf S}
\\ \hline
\end{array}
\end{equation}
Accordingly, the $\bar D^0 \bar D^0$ covalent molecule does not exist.

\item{$\bar D[q_1 \bar c_2]$--$\bar D[q_3 \bar c_4]$.} After including the isospin symmetry, exchange can occur between the $up$ and $down$ quarks. Let us exchange $q_1$ from the first $\bar D$ meson with $q_3$ from the second $\bar D$ meson. As discussed above, they have the symmetric spin structure, so they can be totally antisymmetric as long as their flavor structure is antisymmetric:
\begin{equation}\nonumber
\begin{array}{cccccc}
\hline
                           & {\rm color} & {\rm flavor} & {\rm spin} & {\rm orbital} & {\rm total}
\\ \hline
q_1 \leftrightarrow q_3    & {\bf S}     & {\bf A}      & {\bf S}    & {\bf S}       & {\bf A}
\\ \hline
\end{array}
\end{equation}
However, this configuration still vanishes, {\it i.e.},
\begin{equation}
| \bar D \bar D; I=0 \rangle = \bar D^0 D^- - D^- \bar D^0 = 0 \, ,
\end{equation}
so there is no $\bar D \bar D$ covalent molecule, neither with $I=0$ nor with $I=1$.

\item{$\bar D[q_1 \bar c_2]$--$\bar D^{*}[q_3 \bar c_4]$.} Let us exchange $q_1$ from the $\bar D$ meson with $q_3$ from the $\bar D^{*}$ meson. In this case $q_1$ and $q_3$ do not need to spin in the same direction, since the $\bar D \bar D^{*}$ molecule can transform into the $\bar D^{*} \bar D$ molecule with the exchange of these two light quarks. Accordingly, there are two possible configurations that satisfy the Pauli principle, either
    \begin{equation}\nonumber
    \begin{array}{cccccc}
    \hline
    {\it strong}                         & {\rm color} & {\rm flavor}      & {\rm spin}      & {\rm orbital}      & {\rm total}
    \\ \hline
    q_1 \leftrightarrow q_3              & {\bf S}     & {\bf A}           & {\bf S}         & {\bf S}            & {\bf A}
    \\ \hline
    \end{array}
    \end{equation}
    or
    \begin{equation}\nonumber
    \begin{array}{cccccc}
    \hline
    {\it weak}                           & {\rm color} & {\rm flavor}      & {\rm spin}      & {\rm orbital}      & {\rm total}
    \\ \hline
    q_1 \leftrightarrow q_3              & {\bf S}     & {\bf S}           & {\bf A}         & {\bf S}            & {\bf A}
    \\ \hline
    \end{array}
    \end{equation}
    As discussed in Ref.~\cite{Chen:2021xlu}, the former configuration with $(I)J^P=(0)1^+$ is more stable than the latter one with $(I)J^P=(1)0^+$. Accordingly, we refer to the former as the ``strong'' bond and the latter as the ``weak'' bond. Additionally, there exists a ``repulsive'' bond due to the exchange of the two $charm$ antiquarks:
    \begin{equation}\nonumber
    \begin{array}{cccccc}
    \hline
    {\it repulsive}                      & {\rm color} & {\rm flavor}      & {\rm spin}      & {\rm orbital}      & {\rm total}
    \\ \hline
    \bar c_2 \leftrightarrow \bar c_4    & {\bf S}     & {\bf S}           & {\bf S}         & {\bf S}            & {\bf S}
    \\ \hline
    \end{array}
    \end{equation}
    Therefore, we need to consider both the ``strong/weak'' bond and the ``repulsive'' bond in order to verify the existence of the $I=0/1$ $\bar D \bar D^*$ covalent molecule. Specifically, as illustrated in Fig.~\ref{fig:bond}(a), the $T_{cc}(3875)$ can be interpreted as the $\bar D \bar D^*$ hadronic covalent molecule with $(I)J^P = (0)1^+$, which contains one strong bond and one repulsive bond.

\item{$\bar D^*[q_1 \bar c_2]$--$\bar D^{*}[q_3 \bar c_4]$.} Similarly, we study the $\bar D^* \bar D^*$ covalent molecule. Our results suggest the possible existence of the $(I)J^P=(0)1^+$ $\bar D^* \bar D^*$ covalent molecule, which contains one strong bond and one repulsive bond. However, the $(I)J^P=(0)0^+$ and $(I)J^P=(0)2^+$ $\bar D^* \bar D^*$ covalent molecules do not exist. Besides, our results suggest the possible existence of the $(I)J^P=(1)0^+$ and $(I)J^P=(1)2^+$ $\bar D^* \bar D^*$ covalent molecules, each of which contains one weak bond and one repulsive bond, while the $(I)J^P=(1)1^+$ $\bar D^* \bar D^*$ covalent molecule does not exist.

\end{itemize}
Summarizing the above discussions, our results suggest the possible existence of the $(I)J^P=(0)1^+$ $\bar D \bar D^*$ covalent molecule and the $(I)J^P=(0)1^+$ $\bar D^* \bar D^*$ covalent molecule, both of which contain the strong covalent bond; our results suggest the possible existence of the $(I)J^P=(1)1^+$ $\bar D \bar D^*$ covalent molecule and the $(I)J^P=(1)0/2^+$ $\bar D^* \bar D^*$ covalent molecules, which contain the weak covalent bond.

\subsection{Deuteron}
\label{sec:deuteron}

%
\begin{table*}[htb]
\begin{center}
\renewcommand{\arraystretch}{1.8}
\caption{Binding energies of some possible hadronic covalent molecules, estimated within our toy model through the simplified formula $B = N_S S + N_W W + N_\Lambda \Lambda - N_R R - N \epsilon$, with $S \sim 23$~MeV, $W \sim 17$~MeV, $\Lambda \sim 18$~MeV, $R \sim 14$~MeV, and $\epsilon \sim 4$~MeV. The spin effects are not taken into account in this formula. For brevity, we denote \(D\) and \(D^*\) collectively as \(D^{(*)}\), and \(\Sigma_c\) and \(\Sigma_c^*\) collectively as \(\Sigma_c^{(*)}\), and so on.}
\begin{tabular}{c | c || c | c}
\hline\hline
~~~Covalent molecules~~~ & ~~~Binding energies~~~ & ~~~Covalent molecules~~~ & ~~~Binding energies~~~
\\ \hline \hline
$^2$H, $D^*D^{(*)}/\bar B^*\bar B^{(*)}$                                                      &   1~MeV
&
$D^{*} \bar B^{(*)}/D^{(*)} \bar B^{*}$                                                       &   15~MeV
\\ \hline
$^3$H/$^3$He, $D^*D^{*}D^{(*)}/\bar B^*\bar B^{*}\bar B^{(*)}$                                &   9~MeV
&
$D^{*}D^{*}\bar B^{(*)}/D^{(*)}\bar B^{*}\bar B^{*}/\cdots$                                   &   37~MeV
\\ \hline
\multirow{2}{*}{$^4$He, $D^*D^*D^{*}D^{(*)}/\bar B^*\bar B^*\bar B^{*}\bar B^{(*)}$}          &   \multirow{2}{*}{26~MeV}
&
$D^*D^{*}D^{*}\bar B^{(*)}/D^{(*)}\bar B^{*}\bar B^{*}\bar B^*/\cdots$                        &   68~MeV
\\ \cline{3-4}
& &
$D^{*}D^{*}\bar B^{*}\bar B^{(*)}/D^{(*)}D^{*}\bar B^{*}\bar B^{*}/\cdots$                    &   82~MeV
\\ \hline\hline
$\Sigma_c^{(*)}\Sigma_c^{(*)}/\Sigma_b^{(*)}\Sigma_b^{(*)}$                                   &   24~MeV
&
$\Sigma_c^{(*)}\Sigma_b^{(*)}$                                                                &   38~MeV
\\ \hline\hline
\multicolumn{2}{c||}{} &
$\bar D^{(*)} \Sigma_c^{(*)}/\bar D^{(*)} \Sigma_b^{(*)}/B^{(*)} \Sigma_c^{(*)}/B^{(*)} \Sigma_b^{(*)}$ & 15~MeV
\\ \cline{3-4}
\multicolumn{2}{c||}{} &
$\bar D^{*}\bar D^{(*)}\Sigma_c^{(*)}/\bar D^{*}\bar D^{(*)}\Sigma_b^{(*)}/\cdots$            &   37~MeV
\\ \hline\hline
& &
$D^{*} \bar B^{(*)}_s/D^{(*)} \bar B^{*}_s$                                                   &   10~MeV
\\ \hline
$^3_\Lambda$H, $D^*D^{*}D_s^{(*)}/D^{(*)}D^{*}D_s^{*}$                                        &   1~MeV
&
$D^{*}D^{*}\bar B_s^{(*)}/D^{(*)}D^{*}\bar B_s^{*}/\cdots$                                    &   15~MeV
\\ \hline
$^4_\Lambda$H/$^4_\Lambda$He, $D^*D^{*}D^{*}D_s^{(*)}/D^{(*)}D^{*}D^{*}D_s^*$                 &   13~MeV
&
$D^*D^{*}D^{*}\bar B_s^{(*)}/D^{(*)}D^*D^{*}\bar B_s^{*}/\cdots$                              &   41~MeV
\\ \hline
$^5_\Lambda$He, $D^*D^*D^{*}D^{*}D_s^{(*)}/\cdots$                                            &   30~MeV
&
$D^*D^*D^{*}D^{*}\bar B_s^{(*)}/\cdots$                                                       &   58~MeV
\\ \hline
$^{~~6}_{\Lambda\Lambda}$He, $D^*D^*D^{*}D^{*}D_s^{(*)}D_s^{(*)}/\cdots$                      &   34~MeV
&
$D^*D^*D^{*}D^{*}\bar B_s^{(*)}\bar B_s^{(*)}/\cdots$                                         &   90~MeV
\\ \hline\hline
$\Sigma_c^{(*)}\Xi_c^{(\prime*)}/\Sigma_b^{(*)}\Xi_b^{(\prime*)}$                             &   19~MeV
&
$\Sigma_c^{(*)}\Xi_b^{(\prime*)}/\Sigma_b^{(*)}\Xi_c^{(\prime*)}$                             &   33~MeV
\\ \hline
$\Xi_c^{(\prime*)}\Xi_c^{(\prime*)}/\Xi_b^{(\prime*)}\Xi_b^{(\prime*)}$                       &   14~MeV
&
$\Xi_c^{(\prime*)}\Xi_b^{(\prime*)}$                                                          &   28~MeV
\\ \hline
$\Sigma_c^{(*)}\Xi_c^{(\prime*)}\Xi_c^{(\prime*)}/\Sigma_b^{(*)}\Xi_b^{(\prime*)}\Xi_b^{(\prime*)}$                         &   45~MeV
&
$\Sigma_c^{(*)}\Xi_c^{(\prime*)}\Xi_b^{(\prime*)}/\Sigma_b^{(*)}\Xi_c^{(\prime*)}\Xi_c^{(\prime*)}$                         &   73~MeV
\\ \hline\hline
\multicolumn{2}{c||}{} &
$\bar D^{(*)} \Xi_c^{(\prime*)}/\bar D^{(*)} \Xi_b^{(\prime*)}/B^{(*)} \Xi_c^{(\prime*)}/B^{(*)} \Xi_b^{(\prime*)}$         &   15~MeV
\\ \cline{3-4}
\multicolumn{2}{c||}{} &
$\bar D^{*}\bar D^{(*)}\Xi_c^{(\prime*)}/\bar D^{*}\bar D^{(*)}\Xi_b^{(\prime*)}/\cdots$                                    &   32~MeV
\\ \cline{3-4}
\multicolumn{2}{c||}{} &
$\bar D^{*}\bar D^{(*)}\Xi_c^{(\prime*)}\Xi_c^{(\prime*)}/\bar D^{*}\bar D^{(*)}\Xi_b^{(\prime*)}\Xi_b^{(\prime*)}/\cdots$  &   55~MeV
\\ \hline\hline
\end{tabular}
\label{tab:binding}
\end{center}
\end{table*}
%

In this subsection we study nuclei as the hadronic covalent molecules composed of nucleons. We take the $p(roton)/n(eutron)/\Lambda$ as combinations of an $(I)J^P = (0)0^+$ $up$-$down$ quark pair along with an isolated $up/down/strange$ quark, and explore several examples as follows ($q=up/down$ and $s=strange$):
\begin{itemize}

\item{$N[q_1(u_2d_3)_{(0)0^+}]$--$N[q_4(u_5d_6)_{(0)0^+}]$.} Let us exchange $q_1$ and $q_4$ in order to share them between two nucleons. There are two possible configurations that satisfy the Pauli principle, either
    \begin{equation}\nonumber
    \begin{array}{cccccc}
    \hline
    {\it strong}                         & {\rm color} & {\rm flavor} & {\rm spin} & {\rm orbital} & {\rm total}
    \\ \hline
    q_1 \leftrightarrow q_4              & {\bf S}     & {\bf A}      & {\bf S}    & {\bf S}       & {\bf A}
    \\ \hline
    \end{array}
    \end{equation}
    or
    \begin{equation}\nonumber
    \begin{array}{cccccc}
    \hline
    {\it weak}                           & {\rm color} & {\rm flavor} & {\rm spin} & {\rm orbital} & {\rm total}
    \\ \hline
    q_1 \leftrightarrow q_4              & {\bf S}     & {\bf S}      & {\bf A}    & {\bf S}       & {\bf A}
    \\ \hline
    \end{array}
    \end{equation}
    Similar to the $\bar D \bar D^*$ covalent molecule, the former strong bond is more stable than the latter weak bond, and moreover, there exists a repulsive bond due to the two $(I)J^P = (0)0^+$ $up$-$down$ quark pairs:
    \begin{equation}\nonumber
    \begin{array}{cccccc}
    \hline
    {\it repulsive}                      & {\rm color} & {\rm flavor} & {\rm spin} & {\rm orbital} & {\rm total}
    \\ \hline
    u_2 \leftrightarrow u_5              & {\bf S}     & {\bf S}      & {\bf S}    & {\bf S}       & {\bf S}
    \\
    d_3 \leftrightarrow d_6              & {\bf S}     & {\bf S}      & {\bf S}    & {\bf S}       & {\bf S}
    \\ \hline
    \end{array}
    \end{equation}
    Therefore, we need to consider both the strong/weak bond and the repulsive bond in order to verify the existence of the $I=0/1$ $NN$ covalent molecule. Specifically, the deuteron can be interpreted as the $proton$--$neutron$ hadronic covalent molecule with $(I)J^P = (0)1^+$, which contains one strong bond and one repulsive bond.

\item{$^4$He[$p(u_1)p(u_2)n(d_3)n(d_4)$].} It is interesting to study how many light quarks can be shared at most in the lowest orbit. We find that the $^4$He nucleus can accommodate four light $up/down$ quarks with the configuration of $(I)J^P = (0)0^+$, satisfying that any two of these four quarks are totally antisymmetric so that obey the Pauli principle:
\begin{equation}\nonumber
\begin{array}{cccccc}
\hline
                           & {\rm color} & {\rm flavor} & {\rm spin} & {\rm orbital} & {\rm total}
\\ \hline
u_1 \leftrightarrow u_2    & {\bf S}     & {\bf S}      & {\bf A}    & {\bf S}       & {\bf A}
\\
u_1 \leftrightarrow d_3    & {\bf S}     & {\bf A}      & {\bf S}    & {\bf S}       & {\bf A}
\\
u_1 \leftrightarrow d_4    & {\bf S}     & {\bf A}      & {\bf S}    & {\bf S}       & {\bf A}
\\
u_2 \leftrightarrow d_3    & {\bf S}     & {\bf A}      & {\bf S}    & {\bf S}       & {\bf A}
\\
u_2 \leftrightarrow d_4    & {\bf S}     & {\bf A}      & {\bf S}    & {\bf S}       & {\bf A}
\\
d_3 \leftrightarrow d_4    & {\bf S}     & {\bf S}      & {\bf A}    & {\bf S}       & {\bf A}
\\ \hline
\end{array}
\end{equation}
This makes the $^4$He nucleus quite stable.

\item{$^3_\Lambda$H$[p(u_1)n(d_2)\Lambda(s_3)]$.} Let us exchange $u_1$ from the $proton$, $d_2$ from the $neutron$, and $s_3$ from the $\Lambda$. It is possible for any two of them to be totally antisymmetric, either
    \begin{equation}\nonumber
    \begin{array}{cccccc}
    \hline
                            & {\rm color} & {\rm flavor} & {\rm spin} & {\rm orbital} & {\rm total}
    \\ \hline
    u_1 \leftrightarrow d_2 & {\bf S}     & {\bf A}      & {\bf S}    & {\bf S}       & {\bf A}
    \\
    d_2 \leftrightarrow s_3 & {\bf S}     & {\bf A}      & {\bf S}    & {\bf S}       & {\bf A}
    \\
    s_3 \leftrightarrow u_1 & {\bf S}     & {\bf S}      & {\bf A}    & {\bf S}       & {\bf A}
    \\ \hline
    \end{array}
    \end{equation}
    or
    \begin{equation}\nonumber
    \begin{array}{cccccc}
    \hline
                            & {\rm color} & {\rm flavor} & {\rm spin} & {\rm orbital} & {\rm total}
    \\ \hline
    u_1 \leftrightarrow d_2 & {\bf S}     & {\bf A}      & {\bf S}    & {\bf S}       & {\bf A}
    \\
    d_2 \leftrightarrow s_3 & {\bf S}     & {\bf S}      & {\bf A}    & {\bf S}       & {\bf A}
    \\
    s_3 \leftrightarrow u_1 & {\bf S}     & {\bf A}      & {\bf S}    & {\bf S}       & {\bf A}
    \\ \hline
    \end{array}
    \end{equation}
    We use the strong/weak $qq$ bond to refer to the strong/weak bond composed of two light $up/down$ quarks, and the strong/weak $qs$ bond to refer to the strong/weak bond composed of one light $up/down$ quark and one light $strange$ quark. It seems that the $^3_\Lambda$H nucleus contains one strong $qq$ bond, one strong $qs$ bond, one weak $qs$ bond, and three repulsive bonds. However, the weak $qs$ bond may be even weaker than the repulsive bond, so it is possible that there is only one strong $qq$ bond, one strong $qs$ bond, and two repulsive bonds in this system.

\end{itemize}

\subsection{A toy model}
\label{sec:model}

Many possible hadronic covalent molecules can exist, in addition to the $T_{cc}(3875)$ and the deuteron investigated in the previous subsections. A toy model was developed in Ref.~\cite{Chen:2021xlu} to estimate their binding energies ($B$), and in this subsection we update it as
\begin{equation}
B = N_S S + N_W W + N_\Lambda \Lambda - N_R R - N \epsilon \, ,
\label{eq:formula}
\end{equation}
where $S$ is the energy of the strong $qq$ bond with $N_S$ its number, $W$ is the energy of the weak $qq$ bond with $N_W$ its number, $\Lambda$ is the energy of the strong $qs$ bond with $N_\Lambda$ its number, $R$ is the energy of the repulsive bond with $N_R$ its number, and $\epsilon$ is the residual energy of each component hadron with $N$ the number of components. The weak $qs$ bond is not taken into account. Note that the term $N_W W$ was neglected in Ref.~\cite{Chen:2021xlu}, while the term to describe the spin effects is neglected here, since we do not consider the spin effects in the present study. Assuming the $P_c/P_{cs}/T_{cc}$ as possible hadronic covalent molecules, we estimate these parameters to be
\begin{eqnarray}
\nonumber S &\sim& 23~{\rm MeV} \, ,
\\ \nonumber W &\sim& 17~{\rm MeV} \, ,
\\ \Lambda &\sim& 18~{\rm MeV} \, ,
\\ \nonumber R &\sim& 14~{\rm MeV} \, ,
\\ \nonumber \epsilon &\sim& 4~{\rm MeV} \, .
\end{eqnarray}
We use the above toy model to estimate the binding energies of some possible hadronic covalent molecules, and the obtained results are summarized in Table~\ref{tab:binding}.

\section{Creation bond}
\label{sec:creation}

The hadronic covalent bond studied in the previous section cannot explain the hadronic molecule composed of one $D^{(*)}$ meson and one $\bar D^{(*)}$ meson, which will be explored in this section. To address this, we extend the covalent bond by incorporating not only the light quark-antiquark pair from the $D^{(*)}$ and $\bar D^{(*)}$ mesons, but also two sea quark-antiquark pairs from the vacuum, as illustrated in Fig.~\ref{fig:bond}(b).

We use $\bar q_1 q_2$ to denote the light quark-antiquark pair from the $D^{(*)}$ and $\bar D^{(*)}$ mesons, where $\bar q_1$ is from the $D^{(*)}$ meson and $q_2$ is from the $\bar D^{(*)}$ meson. We use $\bar q_3 q_4$ and $\bar q_5 q_6$ to denote the two sea quark-antiquark pairs from the vacuum. According to the Lagrangian $\mathcal{L} = \bar q i \gamma^\mu D_\mu q$ with $D_\mu = \partial_\mu + i g_s A_\mu$, we assume their quantum numbers to be both
\begin{equation}
[^3S_1]~:~I=0 \, , \, S=1 \, , \, L=0 \, , \, J=1 \, ,
\end{equation}
and their total quantum numbers to be
\begin{equation}
[^3S_1]\otimes[^3S_1]~:~I=0 \, , \, S=0 \, , \, L=0 \, , \, J=0 \, .
\end{equation}
In addition, two $^3P_0$ quark-antiquark pairs are also possible, but they would result in a higher energy configuration. Moreover, we require that the three antiquarks $\bar q_{1,3,5}$, which orbit around the $charm$ quark in the $D^{(*)}$ meson, have the same color and so the symmetric color structure; the three quarks $q_{2,4,6}$, which orbit around the $charm$ antiquark in the $\bar D^{(*)}$ meson, also have the same color and so the symmetric color structure; the two antiquarks $\bar q_3$ and $\bar q_5$ form a strong bond, while the two quarks $q_4$ and $q_6$ form another strong bond. There are numerous sea quark-antiquark pairs and sea gluons in the QCD vacuum, making it possible for these assumptions to be satisfied.

Based on the above assumptions, we find that the $\bar q_1 q_2$ pair with the quantum numbers
\begin{equation}
I=1 \, , \, S=0 \, , \, L=0 \, , \, J=0 \, ,
\label{eq:1000}
\end{equation}
is capable of forming the following configuration:
\begin{equation}\nonumber
\renewcommand{\arraystretch}{1.1}
\begin{array}{cccccc}
\hline\hline
[I_1S_0L_0J_0]                       & {\rm color}          & {\rm flavor} & {\rm spin} & {\rm orbital} & {\rm total}
\\ \hline\hline
\bar q_1 \leftrightarrow q_2         & \mathbf{1_C}         & {\bf S}      & {\bf A}    & {\bf S}       & [^1S_0]
\\ \hline
\bar q_1 \leftrightarrow \bar q_3    & {\bf S}              & {\bf S}      & {\bf A}    & {\bf S}       & {\bf A}~~(weak)~
\\
\bar q_1 \leftrightarrow \bar q_5    & {\bf S}              & {\bf A}      & {\bf S}    & {\bf S}       & {\bf A}~(strong)
\\
\bar q_3 \leftrightarrow \bar q_5    & {\bf S}              & {\bf A}      & {\bf S}    & {\bf S}       & {\bf A}~(strong)
\\
q_2 \leftrightarrow q_4              & {\bf S}              & {\bf A}      & {\bf S}    & {\bf S}       & {\bf A}~(strong)
\\
q_2 \leftrightarrow q_6              & {\bf S}              & {\bf S}      & {\bf A}    & {\bf S}       & {\bf A}~~(weak)~
\\
q_4 \leftrightarrow q_6              & {\bf S}              & {\bf A}      & {\bf S}    & {\bf S}       & {\bf A}~(strong)
\\ \hline
\bar q_3 \leftrightarrow q_4         & \mathbf{8_C}         & {\bf A}      & {\bf S}    & {\bf S}       & [^3S_1]
\\
\bar q_3 \leftrightarrow q_6         & \mathbf{1_C/8_C}     & {\bf S}      & {\bf A}    & {\bf S}       & [\pi(?)]
\\
\bar q_5 \leftrightarrow q_4         & \mathbf{1_C/8_C}     & {\bf S}      & {\bf A}    & {\bf S}       & [\pi(?)]
\\
\bar q_5 \leftrightarrow q_6         & \mathbf{8_C}         & {\bf A}      & {\bf S}    & {\bf S}       & [^3S_1]
\\ \hline
\bar q_1 \leftrightarrow q_4         & -                    & {\bf A/S}    & {\bf S}    & {\bf S}       & -
\\
\bar q_1 \leftrightarrow q_6         & -                    & {\bf S}      & {\bf S}    & {\bf S}       & -
\\
\bar q_3 \leftrightarrow q_2         & -                    & {\bf S}      & {\bf S}    & {\bf S}       & -
\\
\bar q_5 \leftrightarrow q_2         & -                    & {\bf A/S}    & {\bf S}    & {\bf S}       & -
\\ \hline\hline
\end{array}
\end{equation}
We refer to this configuration as the hadronic creation bond due to the activity of sea quark-antiquark pairs from the vacuum, thus being closely related to the QCD confinement. It contains four strong bonds ($\bar q_1\bar q_5$, $\bar q_3\bar q_5$, $q_2q_4$, and $q_4q_6$) and two weak bonds ($\bar q_1\bar q_3$ and $q_2q_6$), which attract the $D^{(*)}$ and $\bar D^{(*)}$ mesons to potentially form a hadronic confined molecule that behaves as a resonance lying above the $D^{(*)} \bar D^{(*)}$ threshold. Additionally, the two color-octet sea quark-antiquark pairs from the vacuum, $\bar q_3 q_4$ and $\bar q_5 q_6$ both with $I=0$ and $J=1$, couple strongly to sea gluons, so their quantity is tremendous; moreover, they can be recombined into the two color-singlet quark-antiquark pairs, $\bar q_3 q_6$ and $\bar q_5 q_4$ both with $I=1$ and $J=0$, which couple strongly to the pseudoscalar mesons $\pi$ and so capable of providing long-range attraction.

Besides, the $\bar q_1 q_2$ pair with the quantum numbers
\begin{equation}
I=0 \, , \, S=1 \, , \, L=0 \, , \, J=1 \, ,
\label{eq:0101}
\end{equation}
can form another similar configuration:
\begin{equation}\nonumber
\renewcommand{\arraystretch}{1.1}
\begin{array}{cccccc}
\hline\hline
[I_0S_1L_0J_1]                       & {\rm color}          & {\rm flavor} & {\rm spin} & {\rm orbital} & {\rm total}
\\ \hline\hline
\bar q_1 \leftrightarrow q_2         & \mathbf{1_C}         & {\bf A}      & {\bf S}    & {\bf S}       & [^3S_1]
\\ \hline
\bar q_1 \leftrightarrow \bar q_3    & {\bf S}              & {\bf S}      & {\bf A}    & {\bf S}       & {\bf A}~~(weak)~
\\
\bar q_1 \leftrightarrow \bar q_5    & {\bf S}              & {\bf A}      & {\bf S}    & {\bf S}       & {\bf A}~(strong)
\\
\bar q_3 \leftrightarrow \bar q_5    & {\bf S}              & {\bf A}      & {\bf S}    & {\bf S}       & {\bf A}~(strong)
\\
q_2 \leftrightarrow q_4              & {\bf S}              & {\bf S}      & {\bf A}    & {\bf S}       & {\bf A}~~(weak)~
\\
q_2 \leftrightarrow q_6              & {\bf S}              & {\bf A}      & {\bf S}    & {\bf S}       & {\bf A}~(strong)
\\
q_4 \leftrightarrow q_6              & {\bf S}              & {\bf A}      & {\bf S}    & {\bf S}       & {\bf A}~(strong)
\\ \hline
\bar q_3 \leftrightarrow q_4         & \mathbf{8_C}         & {\bf A}      & {\bf S}    & {\bf S}       & [^3S_1]
\\
\bar q_3 \leftrightarrow q_6         & \mathbf{1_C/8_C}     & {\bf S}      & {\bf A}    & {\bf S}       & [\pi(?)]
\\
\bar q_5 \leftrightarrow q_4         & \mathbf{1_C/8_C}     & {\bf S}      & {\bf A}    & {\bf S}       & [\pi(?)]
\\
\bar q_5 \leftrightarrow q_6         & \mathbf{8_C}         & {\bf A}      & {\bf S}    & {\bf S}       & [^3S_1]
\\ \hline
\bar q_1 \leftrightarrow q_4         & -                    & {\bf A}      & {\bf A/S}  & {\bf S}       & -
\\
\bar q_1 \leftrightarrow q_6         & -                    & {\bf S}      & {\bf S}    & {\bf S}       & -
\\
\bar q_3 \leftrightarrow q_2         & -                    & {\bf A}      & {\bf A/S}  & {\bf S}       & -
\\
\bar q_5 \leftrightarrow q_2         & -                    & {\bf S}      & {\bf S}    & {\bf S}       & -
\\ \hline\hline
\end{array}
\end{equation}
However, the $\bar q_1 q_2$ pair with the quantum numbers
\begin{equation}
I=0 \, , \, S=0 \, , \, L=0 \, , \, J=0 \, ,
\end{equation}
cannot form this type of configuration, nor can the $\bar q_1 q_2$ pair with the quantum numbers
\begin{equation}
I=1 \, , \, S=1 \, , \, L=0 \, , \, J=1 \, .
\end{equation}
Besides the configurations mentioned above, there are other possible configurations that could form hadronic creation bonds, and thus hadronic confined molecules. We list them in Appendix~\ref{app:configuration}, but we shall not investigate them in the present study.

We apply the above binding mechanism to study several examples, as follows:
\begin{itemize}

\item{$D[\bar q_1 c_7]$--$\bar D[q_2 \bar c_8]$.} The $D$ meson cannot employ the weak $\bar q_1\bar q_3/\bar q_1\bar q_5$ bond, and the $\bar D$ meson cannot employ the weak $q_2q_4/q_2q_6$ bond either. Accordingly, there is no $D \bar D$ confined molecule, neither with $I=0$ nor with $I=1$.

\item{$D[\bar q_1 c_7]$--$\bar D^*[q_2 \bar c_8] \oplus D^*[\bar q_1 c_7]$--$\bar D[q_2 \bar c_8]$.} The $D$ meson itself cannot employ the weak $\bar q_1\bar q_3/\bar q_1\bar q_5$ bond, but the $D$ and $D^*$ mesons together can employ the weak $\bar q_1\bar q_3/\bar q_1\bar q_5$ bond, making it possible for the combination of $D\bar D^*$ and $D^*\bar D$ to form a hadronic confined molecule.

    According to the heavy quark spin symmetry~\cite{Isgur:1989vq,Neubert:1993mb,Manohar:2000dt,Guo:2009id,Bondar:2011ev,Hidalgo-Duque:2013pva,Ozpineci:2013zas}, we perform the spin decomposition:
\begin{eqnarray}
&& |  \left( D \bar D^* - D^* \bar D \right)/\sqrt2 ; J^{PC} = 1^{++} \rangle
\label{def:DDs1++}
\\ \nonumber && ~~~~~~~~~~~~ = -  | s_{\bar c_8 c_7} = 1, s_{\bar q_1 q_2} = 1 ; J = 1 \rangle \, ,
\\[2mm] && | \left( D \bar D^* + D^* \bar D \right)/\sqrt2 ; J^{PC} = 1^{+-} \rangle
\label{def:DDs1+-}
\\ \nonumber && ~~~~~~~~~~~~ = +  | s_{\bar c_8 c_7} = 1, s_{\bar q_1 q_2} = 0 ; J = 1 \rangle/\sqrt2
\\ \nonumber && ~~~~~~~~~~~~~~~-  | s_{\bar c_8 c_7} = 0, s_{\bar q_1 q_2} = 1 ; J = 1 \rangle/\sqrt2 \, .
\end{eqnarray}
There are three possibilities:
\begin{enumerate}
\item The state given in Eq.~(\ref{def:DDs1++}) contains the $\bar q_1 q_2$ pair with $s_{\bar q_1 q_2} = 1$, so its isoscalar component can form a creation bond, making it possible to interpret the $X(3872)$ as the $D \bar D^*/D^* \bar D$ confined molecule with $I^GJ^{PC} = 0^+1^{++}$.
\item The state given in Eq.~(\ref{def:DDs1+-}) contains the $\bar q_1 q_2$ pair with $s_{\bar q_1 q_2} = 0$, so its isovector component can form a creation bond, making it possible to interpret the $Z_c(3900)$ as the $D \bar D^*/D^* \bar D$ confined molecule with $I^GJ^{PC} = 1^+1^{+-}$.
\item The state given in Eq.~(\ref{def:DDs1+-}) also contains the $\bar q_1 q_2$ pair with $s_{\bar q_1 q_2} = 1$, suggesting the possible existence of the $D \bar D^*/D^* \bar D$ confined molecule with $I^GJ^{PC} = 0^-1^{+-}$. However, its corresponding $\bar c_8 c_7$ pair has $s_{\bar c_8 c_7} = 0$, which could render this system potentially unstable.
\end{enumerate}

\item{$D^*[\bar q_1 c_7]$--$\bar D^*[q_2 \bar c_8]$.} According to the heavy quark spin symmetry, we perform the spin decomposition:
\begin{eqnarray}
&& | D^* \bar D^* ; J^{PC} = 0^{++} \rangle
\label{def:DsDs0++}
\\ \nonumber && ~~~~~~~~ = - | s_{\bar c_8 c_7} = 0, s_{\bar q_1 q_2} = 0 ; J = 0 \rangle \times {\sqrt3/2}
\\ \nonumber && ~~~~~~~~~~~+ | s_{\bar c_8 c_7} = 1, s_{\bar q_1 q_2} = 1 ; J = 0 \rangle/2 \, ,
\\[2mm] && | D^* \bar D^* ; J^{PC} = 1^{+-} \rangle
\label{def:DsDs1+-}
\\ \nonumber && ~~~~~~~~ = - | s_{\bar c_8 c_7} = 1, s_{\bar q_1 q_2} = 0 ; J = 1 \rangle/\sqrt2
\\ \nonumber && ~~~~~~~~~~~- | s_{\bar c_8 c_7} = 0, s_{\bar q_1 q_2} = 1 ; J = 1 \rangle/\sqrt2 \, ,
\\[2mm] && | D^* \bar D^* ; J^{PC} = 2^{++} \rangle
\label{def:DsDs2++}
\\ \nonumber && ~~~~~~~~ = - | s_{\bar c_8 c_7} = 1, s_{\bar q_1 q_2} = 1 ; J = 2 \rangle \, .
\end{eqnarray}
After requiring $s_{\bar c_8 c_7} = 1$, there are three possibilities remaining:
\begin{enumerate}
\item The state given in Eq.~(\ref{def:DsDs0++}) contains the $\bar q_1 q_2$ pair with $s_{\bar q_1 q_2} = 1$, so its isoscalar component can form a creation bond, suggesting the possible existence of the $D^* \bar D^*$ confined molecule with $I^GJ^{PC} = 0^+0^{++}$.
\item The state given in Eq.~(\ref{def:DsDs1+-}) contains the $\bar q_1 q_2$ pair with $s_{\bar q_1 q_2} = 0$, so its isovector component can form a creation bond, suggesting the possible existence of the $D^* \bar D^*$ confined molecule with $I^GJ^{PC} = 1^+1^{+-}$.
\item The state given in Eq.~(\ref{def:DsDs2++}) contains the $\bar q_1 q_2$ pair with $s_{\bar q_1 q_2} = 1$, so its isoscalar component can form a creation bond, suggesting the possible existence of the $D^* \bar D^*$ confined molecule with $I^GJ^{PC} = 0^+2^{++}$.
\end{enumerate}

\end{itemize}
Summarizing the above discussions, our results suggest the possible existence of the $I^GJ^{PC} = 0^+1^{++}/1^+1^{+-}$ $D \bar D^*/D^* \bar D$ confined molecules and the $I^GJ^{PC} = 0^+0^{++}/1^+1^{+-}/0^+2^{++}$ $D^* \bar D^*$ confined molecules, all of which contain the hadronic creation bond.

\section{Summary and Discussions}
\label{sec:summary}

In this paper we systematically study the $\bar D^{(*)} \bar D^{(*)}$ and $D^{(*)} \bar D^{(*)}$ hadronic molecules. For the $\bar D^{(*)} \bar D^{(*)}$ hadronic molecules, we proposed in Ref.~\cite{Chen:2021xlu} the hadronic covalent bond induced by shared light quarks, and used this binding mechanism to interpret the $T_{cc}(3875)$ as the $\bar D \bar D^*$ covalent molecule with $(I)J^P = (0)1^+$. In this paper we improve upon this binding mechanism and estimate the binding energies of some possible hadronic covalent molecules. The obtained results are summarized in Table~\ref{tab:binding}, based on the hypothesis proposed in Ref.~\cite{Chen:2021xlu} that ``{\it the light-quark-exchange interaction is attractive when the shared light quarks are totally antisymmetric so that obey the Pauli principle}''.

The covalent bond cannot explain the $D^{(*)} \bar D^{(*)}$ hadronic molecules. In this paper we attempt to explain them by incorporating not only the light quark-antiquark pair from the $D^{(*)}$ and $\bar D^{(*)}$ mesons, but also two sea quark-antiquark pairs from the vacuum~\footnote{We also attempted to use only one sea quark-antiquark pair instead of two, but were unable to obtain a reasonable result. However, we still consider this a potential option.}. These three quarks and three antiquarks together can form four strong covalent bonds and two weak covalent bonds, potentially attracting the $D^{(*)}$ and $\bar D^{(*)}$ mesons to form a hadronic confined molecule. Our results suggest the possible existence of the $I^GJ^{PC} = 0^+1^{++}/1^+1^{+-}$ $D \bar D^*/D^* \bar D$ confined molecules and the $I^GJ^{PC} = 0^+0^{++}/1^+1^{+-}/0^+2^{++}$ $D^* \bar D^*$ confined molecules.

Given that the creation of sea quark-antiquark pairs from the vacuum has been considered, their annihilation back into the vacuum should receive equal consideration:
\begin{itemize}

\item The annihilation bond does not affect the isovector $D^{(*)} \bar D^{(*)}$ molecule, but it could potentially destabilize the isoscalar $D^{(*)} \bar D^{(*)}$ molecule unless a relevant charmonium state is nearby (consequently, there would be mixing between the molecular state and the charmonium state, and the mass of this charmonium state would increase). Accordingly, the $I^GJ^{PC} = 0^+0^{++}/0^+2^{++}$ $D^* \bar D^*$ confined molecules might not exist.

\item The annihilation bond could reduce the mass of the molecule due to its mixing with the relevant charmonium state. Accordingly, the mass of the $X(3872)$, interpreted as the $D \bar D^*/D^* \bar D$ confined molecule of $I^GJ^{PC} = 0^+1^{++}$ mixed with the $\chi_{c1}(2P)$ state, becomes smaller than the mass of the $Z_c(3900)$, interpreted as the $D \bar D^*/D^* \bar D$ confined molecule of $I^GJ^{PC} = 1^+1^{+-}$.

\end{itemize}
The hadronic creation and annihilation bonds provide a quasi-static, low-energy platform for studying QCD confinement. In addition to the $D^{(*)} \bar D^{(*)}$ confined molecules, a brief application of these two bonds to the hidden-charm baryonium states is presented in Appendix~\ref{app:baryonium}. We will apply this framework to study additional properties of hadrons in the future, such as the electromagnetic structure of the deuteron.

To end this paper, we list several unique features of our framework, which is based on the hadronic covalent, creation, and annihilation bonds:
\begin{itemize}

\item The hadronic covalent molecule formed by shared light quarks often behaves as a bound state. By contrast, the hadronic confined molecule formed by the shared light quark-antiquark pair often behaves as a resonance lying above the relevant threshold due to the activity of sea quark-antiquark pairs from the vacuum, which might be connected to the behavior of hadrons lying above the thresholds of current quarks.

\item The $D D/\bar B \bar B$ covalent molecules do not exist, neither for $I=0$ nor for $I=1$; the $D \bar D/ B \bar B$ confined molecules possibly do not exist, neither for $I=0$ nor for $I=1$; however, the $\bar D \Sigma_c/\bar D \Sigma_b/B \Sigma_c/B \Sigma_b$ covalent molecules with $(I)J^P = (1/2)1/2^+$ do exist. Note that if the $D \bar D/ B \bar B$ confined molecules exist, we would need to consider the configurations provided in Appendix~\ref{app:configuration} and significantly modify our framework.

\item The binding energies of the $(I)J^P = (0)1^+$ $D\bar B^*/D^* \bar B$ covalent molecules are much larger than those of the $(I)J^P = (0)1^+$ $DD^*/\bar B \bar B^*$ covalent molecules, while the $(I)J^P = (1/2)1/2^+$ $\bar D \Sigma_c/\bar D \Sigma_b/B \Sigma_c/B \Sigma_b$ covalent molecules have similar binding energies.

\end{itemize}

\section*{Acknowledgments}

This project is supported by
the National Natural Science Foundation of China under Grant No.~12075019,
the Jiangsu Provincial Double-Innovation Program under Grant No.~JSSCRC2021488,
and
the Fundamental Research Funds for the Central Universities.

\appendix

\section{Other configurations for the creation bond}
\label{app:configuration}

Besides the configurations provided in Sec.~\ref{sec:creation}, there are other possible configurations that could form hadronic confined molecules. We list them as follows:
\begin{itemize}

\item The $\bar q_1 q_2$ pair with the quantum numbers
\begin{equation}
I=1 \, , \, S=0 \, , \, L=0 \, , \, J=0 \, ,
\end{equation}
can form only one configuration, which has been provided in Sec.~\ref{sec:creation}.

\item Besides the configuration provided in Sec.~\ref{sec:creation}, the $\bar q_1 q_2$ pair with the quantum numbers
\begin{equation}
I=0 \, , \, S=1 \, , \, L=0 \, , \, J=1 \, ,
\end{equation}
can form another configuration:
\begin{equation}\nonumber
\renewcommand{\arraystretch}{1.2}
\begin{array}{cccccc}
\hline\hline
[I_0S_1L_0J_1]                       & {\rm color}          & {\rm flavor} & {\rm spin} & {\rm orbital} & {\rm total}
\\ \hline\hline
\bar q_1 \leftrightarrow q_2         & \mathbf{1_C}         & {\bf A}      & {\bf S}    & {\bf S}       & [^3S_1]
\\ \hline
\bar q_1 \leftrightarrow \bar q_3    & {\bf S}              & {\bf A}      & {\bf S}    & {\bf S}       & {\bf A}~(strong)
\\
\bar q_1 \leftrightarrow \bar q_5    & {\bf S}              & {\bf A}      & {\bf S}    & {\bf S}       & {\bf A}~(strong)
\\
\bar q_3 \leftrightarrow \bar q_5    & {\bf S}              & {\bf S}      & {\bf A}    & {\bf S}       & {\bf A}~~(weak)~
\\
q_2 \leftrightarrow q_4              & {\bf S}              & {\bf A}      & {\bf S}    & {\bf S}       & {\bf A}~(strong)
\\
q_2 \leftrightarrow q_6              & {\bf S}              & {\bf A}      & {\bf S}    & {\bf S}       & {\bf A}~(strong)
\\
q_4 \leftrightarrow q_6              & {\bf S}              & {\bf S}      & {\bf A}    & {\bf S}       & {\bf A}~~(weak)~
\\ \hline
\bar q_3 \leftrightarrow q_4         & -                    & {\bf A}      & {\bf A/S}  & {\bf S}       & -
\\
\bar q_3 \leftrightarrow q_6         & -                    & {\bf A}      & {\bf A/S}  & {\bf S}       & -
\\
\bar q_5 \leftrightarrow q_4         & -                    & {\bf A}      & {\bf A/S}  & {\bf S}       & -
\\
\bar q_5 \leftrightarrow q_6         & -                    & {\bf A}      & {\bf A/S}  & {\bf S}       & -
\\ \hline
\bar q_1 \leftrightarrow q_4         & -                    & {\bf S}      & {\bf A/S}  & {\bf S}       & -
\\
\bar q_1 \leftrightarrow q_6         & -                    & {\bf S}      & {\bf A/S}  & {\bf S}       & -
\\
\bar q_3 \leftrightarrow q_2         & -                    & {\bf S}      & {\bf A/S}  & {\bf S}       & -
\\
\bar q_5 \leftrightarrow q_2         & -                    & {\bf S}      & {\bf A/S}  & {\bf S}       & -
\\ \hline\hline
\end{array}
\end{equation}

\item The $\bar q_1 q_2$ pair with the quantum numbers
\begin{equation}
I=0 \, , \, S=0 \, , \, L=0 \, , \, J=0 \, .
\end{equation}
is capable of forming two configurations, either
\begin{equation}\nonumber
\renewcommand{\arraystretch}{1.2}
\begin{array}{cccccc}
\hline\hline
[I_0S_0L_0J_0]                       & {\rm color}          & {\rm flavor} & {\rm spin} & {\rm orbital} & {\rm total}
\\ \hline\hline
\bar q_1 \leftrightarrow q_2         & \mathbf{1_C}         & {\bf A}      & {\bf A}    & {\bf S}       & [^1S_0]
\\ \hline
\bar q_1 \leftrightarrow \bar q_3    & {\bf S}              & {\bf S}      & {\bf A}    & {\bf S}       & {\bf A}~~(weak)~
\\
\bar q_1 \leftrightarrow \bar q_5    & {\bf S}              & {\bf A}      & {\bf S}    & {\bf S}       & {\bf A}~(strong)
\\
\bar q_3 \leftrightarrow \bar q_5    & {\bf S}              & {\bf A}      & {\bf S}    & {\bf S}       & {\bf A}~(strong)
\\
q_2 \leftrightarrow q_4              & {\bf S}              & {\bf S}      & {\bf A}    & {\bf S}       & {\bf A}~~(weak)~
\\
q_2 \leftrightarrow q_6              & {\bf S}              & {\bf A}      & {\bf S}    & {\bf S}       & {\bf A}~(strong)
\\
q_4 \leftrightarrow q_6              & {\bf S}              & {\bf A}      & {\bf S}    & {\bf S}       & {\bf A}~(strong)
\\ \hline
\bar q_3 \leftrightarrow q_4         & -                    & {\bf A}      & {\bf A}    & {\bf S}       & -
\\
\bar q_3 \leftrightarrow q_6         & -                    & {\bf S}      & {\bf S}    & {\bf S}       & -
\\
\bar q_5 \leftrightarrow q_4         & -                    & {\bf S}      & {\bf S}    & {\bf S}       & -
\\
\bar q_5 \leftrightarrow q_6         & -                    & {\bf A}      & {\bf A}    & {\bf S}       & -
\\ \hline
\bar q_1 \leftrightarrow q_4         & -                    & {\bf A}      & {\bf S}    & {\bf S}       & -
\\
\bar q_1 \leftrightarrow q_6         & -                    & {\bf S}      & {\bf S}    & {\bf S}       & -
\\
\bar q_3 \leftrightarrow q_2         & -                    & {\bf A}      & {\bf S}    & {\bf S}       & -
\\
\bar q_5 \leftrightarrow q_2         & -                    & {\bf S}      & {\bf S}    & {\bf S}       & -
\\ \hline\hline
\end{array}
\end{equation}
or
\begin{equation}\nonumber
\renewcommand{\arraystretch}{1.2}
\begin{array}{cccccc}
\hline\hline
[I_0S_0L_0J_0]                       & {\rm color}          & {\rm flavor} & {\rm spin} & {\rm orbital} & {\rm total}
\\ \hline\hline
\bar q_1 \leftrightarrow q_2         & \mathbf{1_C}         & {\bf A}      & {\bf A}    & {\bf S}       & [^1S_0]
\\ \hline
\bar q_1 \leftrightarrow \bar q_3    & {\bf S}              & {\bf A}      & {\bf S}    & {\bf S}       & {\bf A}~(strong)
\\
\bar q_1 \leftrightarrow \bar q_5    & {\bf S}              & {\bf A}      & {\bf S}    & {\bf S}       & {\bf A}~(strong)
\\
\bar q_3 \leftrightarrow \bar q_5    & {\bf S}              & {\bf S}      & {\bf A}    & {\bf S}       & {\bf A}~~(weak)~
\\
q_2 \leftrightarrow q_4              & {\bf S}              & {\bf A}      & {\bf S}    & {\bf S}       & {\bf A}~(strong)
\\
q_2 \leftrightarrow q_6              & {\bf S}              & {\bf A}      & {\bf S}    & {\bf S}       & {\bf A}~(strong)
\\
q_4 \leftrightarrow q_6              & {\bf S}              & {\bf S}      & {\bf A}    & {\bf S}       & {\bf A}~~(weak)~
\\ \hline
\bar q_3 \leftrightarrow q_4         & -                    & {\bf A}      & {\bf A/S}  & {\bf S}       & -
\\
\bar q_3 \leftrightarrow q_6         & -                    & {\bf A}      & {\bf A/S}  & {\bf S}       & -
\\
\bar q_5 \leftrightarrow q_4         & -                    & {\bf A}      & {\bf A/S}  & {\bf S}       & -
\\
\bar q_5 \leftrightarrow q_6         & -                    & {\bf A}      & {\bf A/S}  & {\bf S}       & -
\\ \hline
\bar q_1 \leftrightarrow q_4         & -                    & {\bf S}      & {\bf A/S}  & {\bf S}       & -
\\
\bar q_1 \leftrightarrow q_6         & -                    & {\bf S}      & {\bf A/S}  & {\bf S}       & -
\\
\bar q_3 \leftrightarrow q_2         & -                    & {\bf S}      & {\bf A/S}  & {\bf S}       & -
\\
\bar q_5 \leftrightarrow q_2         & -                    & {\bf S}      & {\bf A/S}  & {\bf S}       & -
\\ \hline\hline
\end{array}
\end{equation}

\item The $\bar q_1 q_2$ pair with the quantum numbers
\begin{equation}
I=1 \, , \, S=1 \, , \, L=0 \, , \, J=1 \, ,
\end{equation}
is capable of forming the following configuration:
\begin{equation}\nonumber
\renewcommand{\arraystretch}{1.2}
\begin{array}{cccccc}
\hline\hline
[I_1S_1L_0J_1]                       & {\rm color}          & {\rm flavor} & {\rm spin} & {\rm orbital} & {\rm total}
\\ \hline\hline
\bar q_1 \leftrightarrow q_2         & \mathbf{1_C}         & {\bf S}      & {\bf S}    & {\bf S}       & [^3S_1]
\\ \hline
\bar q_1 \leftrightarrow \bar q_3    & {\bf S}              & {\bf S}      & {\bf A}    & {\bf S}       & {\bf A}~~(weak)~
\\
\bar q_1 \leftrightarrow \bar q_5    & {\bf S}              & {\bf A}      & {\bf S}    & {\bf S}       & {\bf A}~(strong)
\\
\bar q_3 \leftrightarrow \bar q_5    & {\bf S}              & {\bf A}      & {\bf S}    & {\bf S}       & {\bf A}~(strong)
\\
q_2 \leftrightarrow q_4              & {\bf S}              & {\bf S}      & {\bf A}    & {\bf S}       & {\bf A}~~(weak)~
\\
q_2 \leftrightarrow q_6              & {\bf S}              & {\bf A}      & {\bf S}    & {\bf S}       & {\bf A}~(strong)
\\
q_4 \leftrightarrow q_6              & {\bf S}              & {\bf A}      & {\bf S}    & {\bf S}       & {\bf A}~(strong)
\\ \hline
\bar q_3 \leftrightarrow q_4         & -                    & {\bf S}      & {\bf S}    & {\bf S}       & -
\\
\bar q_3 \leftrightarrow q_6         & -                    & {\bf A/S}    & {\bf A}    & {\bf S}       & -
\\
\bar q_5 \leftrightarrow q_4         & -                    & {\bf A/S}    & {\bf A}    & {\bf S}       & -
\\
\bar q_5 \leftrightarrow q_6         & -                    & {\bf S}      & {\bf S}    & {\bf S}       & -
\\ \hline
\bar q_1 \leftrightarrow q_4         & -                    & {\bf S}      & {\bf A/S}  & {\bf S}       & -
\\
\bar q_1 \leftrightarrow q_6         & -                    & {\bf A/S}    & {\bf S}    & {\bf S}       & -
\\
\bar q_3 \leftrightarrow q_2         & -                    & {\bf S}      & {\bf A/S}  & {\bf S}       & -
\\
\bar q_5 \leftrightarrow q_2         & -                    & {\bf A/S}    & {\bf S}    & {\bf S}       & -
\\ \hline\hline
\end{array}
\end{equation}

\end{itemize}
The above configurations might form some hadronic confined molecules. Furthermore, we can incorporate the light $strange$ quark to construct more configurations, which could potentially form additional hadronic confined molecules to explain the $Z_{cs}$ states. All these possibilities will be explored in our future studies.

\section{A brief investigation into the hidden-charm baryonium state}
\label{app:baryonium}

In this appendix we briefly investigate the hidden-charm baryonium state, {\it i.e.}, the hadronic molecule composed of one charmed antibaryon and one charmed baryon, which have the quark contents $\bar q_1 \bar q_7 \bar c_9$ and $q_2 q_8 c_{10}$, respectively. We use $\bar q_3 q_4$ and $\bar q_5 q_6$ to denote the two sea quark-antiquark pairs from the vacuum. We require that the three antiquarks $\bar q_{1,3,5}$ have the same color, and the three quarks $q_{2,4,6}$ also have the same color. Utilizing the hadronic creation and annihilation bonds, we examine several examples as follows:
\begin{itemize}

\item{$\bar \Lambda_c[\bar q_1 \bar q_7 \bar c_9]$--$\Lambda_c[q_2 q_8 c_{10}]$.} Let us exchange the light quark $q_2$ from the $\Lambda_c$ baryon with the sea quark $q_4$ from the vacuum. $q_2$ and $q_8$ inside the $\Lambda_c$ baryon spin in opposite directions, $q_4$ and $q_8$ also need to spin in opposite directions in order to form another $\Lambda_c$ baryon, so $q_2$ and $q_4$ spin in the same direction with the symmetric spin structure. Similarly, $q_2$ and $q_4$ have parallel isospin with the symmetric flavor structure. As a result, the two light quarks $q_2$ and $q_4$ are totally symmetric:
\begin{equation}\nonumber
\begin{array}{cccccc}
\hline
                           & {\rm color} & {\rm flavor} & {\rm spin} & {\rm orbital} & {\rm total}
\\ \hline
q_2 \leftrightarrow q_4    & {\bf S}     & {\bf S}      & {\bf S}    & {\bf S}       & {\bf S}
\\ \hline
\end{array}
\end{equation}
Accordingly, there is no $\Lambda_c \bar \Lambda_c$ confined molecule.

\item{$\bar \Lambda_c[\bar q_1 \bar q_7 \bar c_9]$--$\Sigma_c[q_2 q_8 c_{10}] \oplus \bar \Sigma_c[\bar q_1 \bar q_7 \bar c_9]$--$\Lambda_c[q_2 q_8 c_{10}]$.} As discussed above, the $\Lambda_c$ baryon itself cannot employ a strong bond, but the $\Lambda_c$ and $\Sigma_c$ baryons together can, making it possible for the combination of $\bar \Lambda_c \Sigma_c$ and $\bar \Sigma_c \Lambda_c$ to form a hadronic confined molecule. This state has $I=1$, so we can assume that the $\bar q_1 q_2$ pair also has $I=1$. Since there is no constraint on the spin of this quark-antiquark pair, the following configuration can potentially be formed:
\begin{equation}\nonumber
\renewcommand{\arraystretch}{1.1}
\begin{array}{cccccc}
\hline\hline
[I_1S_0L_0J_0]                       & {\rm color}          & {\rm flavor} & {\rm spin} & {\rm orbital} & {\rm total}
\\ \hline\hline
\bar q_1 \leftrightarrow q_2         & \mathbf{1_C}         & {\bf S}      & {\bf A}    & {\bf S}       & [^1S_0]
\\ \hline
\bar q_1 \leftrightarrow \bar q_3    & {\bf S}              & {\bf S}      & {\bf A}    & {\bf S}       & {\bf A}~~(weak)~
\\
\bar q_1 \leftrightarrow \bar q_5    & {\bf S}              & {\bf A}      & {\bf S}    & {\bf S}       & {\bf A}~(strong)
\\
\bar q_3 \leftrightarrow \bar q_5    & {\bf S}              & {\bf A}      & {\bf S}    & {\bf S}       & {\bf A}~(strong)
\\
q_2 \leftrightarrow q_4              & {\bf S}              & {\bf A}      & {\bf S}    & {\bf S}       & {\bf A}~(strong)
\\
q_2 \leftrightarrow q_6              & {\bf S}              & {\bf S}      & {\bf A}    & {\bf S}       & {\bf A}~~(weak)~
\\
q_4 \leftrightarrow q_6              & {\bf S}              & {\bf A}      & {\bf S}    & {\bf S}       & {\bf A}~(strong)
\\ \hline
\bar q_3 \leftrightarrow q_4         & \mathbf{8_C}         & {\bf A}      & {\bf S}    & {\bf S}       & [^3S_1]
\\
\bar q_3 \leftrightarrow q_6         & \mathbf{1_C/8_C}     & {\bf S}      & {\bf A}    & {\bf S}       & [\pi(?)]
\\
\bar q_5 \leftrightarrow q_4         & \mathbf{1_C/8_C}     & {\bf S}      & {\bf A}    & {\bf S}       & [\pi(?)]
\\
\bar q_5 \leftrightarrow q_6         & \mathbf{8_C}         & {\bf A}      & {\bf S}    & {\bf S}       & [^3S_1]
\\ \hline
\bar q_1 \leftrightarrow q_4         & -                    & {\bf A/S}    & {\bf S}    & {\bf S}       & -
\\
\bar q_1 \leftrightarrow q_6         & -                    & {\bf S}      & {\bf S}    & {\bf S}       & -
\\
\bar q_3 \leftrightarrow q_2         & -                    & {\bf S}      & {\bf S}    & {\bf S}       & -
\\
\bar q_5 \leftrightarrow q_2         & -                    & {\bf A/S}    & {\bf S}    & {\bf S}       & -
\\ \hline\hline
\end{array}
\end{equation}
This configuration is identical to the one formed by the quark-antiquark pair given in Eq.~(\ref{eq:1000}). Its corresponding hadronic creation bond attracts the combination of $\bar \Lambda_c \Sigma_c$ and $\bar \Sigma_c \Lambda_c$ to potentially form a hadronic confined molecule. However, the hadronic annihilation bond may annihilate the $\bar q_7 q_8$ pair depending on its isospin, while leaving the isovector $\bar q_1 q_2$ pair unaffected.

\item{$\bar \Sigma_c[\bar q_1 \bar q_7 \bar c_9]$--$\Sigma_c[q_2 q_8 c_{10}]$.} This molecule can have $I = 0$, $1$, or $2$. Specifically, both the $\bar q_1 q_2$ and $\bar q_7 q_8$ pairs inside the isotensor state have $I=1$. The $\bar q_1 q_2$ pair can form a configuration identical to the one described above, and the hadronic annihilation bond does not affect either pair, suggesting that the isotensor $\bar \Sigma_c \Sigma_c$ confined molecule is likely to exist.

\item{$\bar \Xi_c^\prime[\bar q_1 \bar s_7 \bar c_9]$--$\Xi_c^\prime[q_2 s_8 c_{10}]$.} This molecule can have $I=0$ or $I=1$. Specifically, the $\bar q_1 q_2$ pair inside the isoscalar state has $I=0$. Since there is no constraint on its spin, the following configuration can potentially be formed:
\begin{equation}\nonumber
\renewcommand{\arraystretch}{1.1}
\begin{array}{cccccc}
\hline\hline
[I_0S_1L_0J_1]                       & {\rm color}          & {\rm flavor} & {\rm spin} & {\rm orbital} & {\rm total}
\\ \hline\hline
\bar q_1 \leftrightarrow q_2         & \mathbf{1_C}         & {\bf A}      & {\bf S}    & {\bf S}       & [^3S_1]
\\ \hline
\bar q_1 \leftrightarrow \bar q_3    & {\bf S}              & {\bf S}      & {\bf A}    & {\bf S}       & {\bf A}~~(weak)~
\\
\bar q_1 \leftrightarrow \bar q_5    & {\bf S}              & {\bf A}      & {\bf S}    & {\bf S}       & {\bf A}~(strong)
\\
\bar q_3 \leftrightarrow \bar q_5    & {\bf S}              & {\bf A}      & {\bf S}    & {\bf S}       & {\bf A}~(strong)
\\
q_2 \leftrightarrow q_4              & {\bf S}              & {\bf S}      & {\bf A}    & {\bf S}       & {\bf A}~~(weak)~
\\
q_2 \leftrightarrow q_6              & {\bf S}              & {\bf A}      & {\bf S}    & {\bf S}       & {\bf A}~(strong)
\\
q_4 \leftrightarrow q_6              & {\bf S}              & {\bf A}      & {\bf S}    & {\bf S}       & {\bf A}~(strong)
\\ \hline
\bar q_3 \leftrightarrow q_4         & \mathbf{8_C}         & {\bf A}      & {\bf S}    & {\bf S}       & [^3S_1]
\\
\bar q_3 \leftrightarrow q_6         & \mathbf{1_C/8_C}     & {\bf S}      & {\bf A}    & {\bf S}       & [\pi(?)]
\\
\bar q_5 \leftrightarrow q_4         & \mathbf{1_C/8_C}     & {\bf S}      & {\bf A}    & {\bf S}       & [\pi(?)]
\\
\bar q_5 \leftrightarrow q_6         & \mathbf{8_C}         & {\bf A}      & {\bf S}    & {\bf S}       & [^3S_1]
\\ \hline
\bar q_1 \leftrightarrow q_4         & -                    & {\bf A}      & {\bf A/S}  & {\bf S}       & -
\\
\bar q_1 \leftrightarrow q_6         & -                    & {\bf S}      & {\bf S}    & {\bf S}       & -
\\
\bar q_3 \leftrightarrow q_2         & -                    & {\bf A}      & {\bf A/S}  & {\bf S}       & -
\\
\bar q_5 \leftrightarrow q_2         & -                    & {\bf S}      & {\bf S}    & {\bf S}       & -
\\ \hline\hline
\end{array}
\end{equation}
This configuration is identical to the one formed by the quark-antiquark pair given in Eq.~(\ref{eq:0101}). Its corresponding hadronic creation bond attracts the $\bar \Xi_c^\prime$ and $\Xi_c^\prime$ baryons to potentially form a hadronic confined molecule. However, the hadronic annihilation bond can annihilate the $\bar q_1 q_2$ pair, potentially destabilizing this molecule.

\end{itemize}
Summarizing the above discussions, our results suggest the possible existence of the $\bar \Lambda_c \Sigma_c/\bar \Sigma_c \Lambda_c$, $\bar \Sigma_c \Sigma_c$, and $\bar \Xi_c^\prime \Xi_c^\prime$ confined molecules, along with many others. These states may behave as resonances lying above the relevant thresholds due to the activity of sea quark-antiquark pairs from the vacuum, possibly linked to the behavior of hadrons lying above the thresholds of current quarks. These hadronic molecules are formed by the hadronic creation bond proposed in the present study, while the annihilation bond also proposed in the present study may destabilize them. A detailed analysis will be presented in our future work.

\newpage
\newpage
\newpage

\bibliographystyle{elsarticle-num}
\bibliography{ref}

\end{document}